\documentclass[aps,prx,showpacs,twocolumn,reprint,superscriptaddress]{revtex4-2}

\usepackage{qcircuit}
\usepackage[dvips]{graphicx}
\usepackage{amsmath,amssymb,amsthm,mathrsfs,amsfonts,dsfont,mathtools}
\usepackage{epsfig}
\usepackage{braket}
\usepackage{hyperref}
\usepackage{bm}
\usepackage{enumerate}
\usepackage{color}
\usepackage{graphicx}
\usepackage{enumitem}
\usepackage[capitalize]{cleveref}
\usepackage[normalem]{ulem}

\newcommand{\sign}{\text{sign}}
\newcommand{\tr}{\mathrm{Tr}}

\newcommand{\im}{\mathrm{Im}}
\newcommand{\re}{\mathrm{Re}}
\newcommand{\var}{\mathrm{Var}}

\usepackage{amsthm}

\newtheorem{statement}{Statement}

\Crefname{theorem}{Theorem}{Theorems}
\theoremstyle{remark}

\newcommand{\udes}{\mathcal{U}_{desired}}
\newcommand{\ul}{\mathcal{U}_l}
\newcommand{\ndct}{N_{dict}}

\begin{document}	
	
\title{Sparse Probabilistic Synthesis of Quantum Operations}

\author{B\'alint Koczor}
\email{koczor@maths.ox.ac.uk}
\affiliation{Mathematical Institute, University of Oxford, Woodstock Road, Oxford OX2 6GG, United Kingdom}
\affiliation{Department of Materials, University of Oxford, Parks Road, Oxford OX1 3PH, United Kingdom}

\begin{abstract}
Successful implementations of quantum technologies require protocols and algorithms that use as few quantum resources as possible. However, many important quantum operations, such as continuous rotation gates in quantum computing or broadband pulses in NMR or MRI applications, can only be implemented approximately using finite quantum resources. This work develops an approach that enables---at the cost of a modestly increased measurement repetition rate---exact implementations on average. One proceeds by first building a library of a large number of different approximations to the desired gate operation; by randomly selecting these operations according to a pre-optimised probability distribution, one can on average implement the desired operation with a rigorously controllable approximation error. The approach relies on sophisticated tools from convex optimisation to efficiently find optimal probability distributions. A diverse spectrum of applications are demonstrated as (a) exactly synthesising rotations in fault-tolerant quantum computers using only low T-count circuits and (b) synthesising broadband and band-selective pulses of superior performance in quantum optimal control with (c) further applications in NMR or MRI. The approach is very general and a broad spectrum of practical applications in quantum technologies are explicitly demonstrated.
\end{abstract}

\maketitle

\section{Introduction}
With the dawn of quantum computing~\cite{aruteQuantumSupremacyUsing2019, wuStrongQuantumComputational2021, zhongPhaseProgrammableGaussianBoson2021, bluvstein_logical_2023, kim_evidence_2023, acharya_suppressing_2023} and other advanced quantum technologies~\cite{acin2018quantum}, 
the central experimental/engineering task remains to increase coherence times.
It is, however, equally important to find new theoretical tools and protocols that reduce
the required quantum resources, such as reducing stringent coherence-time requirements~\cite{cerezoVariationalQuantumAlgorithms2021a,endoHybridQuantumClassicalAlgorithms2021,bharti2021noisy,van2021measurement,koczor2020quantumAnalytic,PhysRevA.106.062416}. 

Here we develop an approach that allows us to implement an ideal, desired quantum operation $\udes$
that is a central resource in quantum technologies but would otherwise not be possible to directly implement,
e.g., due to T-count limitations in quantum computing, finite rotation-angle resolution in quantum simulators 
or due to finite offset effects in quantum optimum control, quantum metrology, NMR or MRI applications.
The present approach overcomes these limitations by approximating the ideal operation $\udes$ as a linear
combination of a large number of operations $\ul$ that the hardware can natively implement.
We start with an offline pre-processing step whereby  we carefully construct a library of gate operations $\ul$
and efficiently find a sparse quasiprobability
decomposition $\gamma_l$  through the use of established tools in convex optimisation. 
The online, experimental implementation then proceeds by randomly choosing the operations $\ul$ according to
probabilities determined by $\gamma_l$. 
This protocol can on average implement the desired quantum operation $\udes$ through a rigorously controllable,
arbitrarily small approximation error. 

The present approach resembles to certain quantum error mitigation techniques which, however, address a distinct problem
as they mitigate the effect of hardware imperfections~\cite{cai2022quantum,PhysRevX.11.031057}.
In fact, we will first assume gate operations are noise-free -- while we discuss below that indeed the approach can naturally be combined
with quantum error mitigation protocols.
Furthermore, Probabilistic Error Cancellation~\cite{practical_QEM,gambetta_error_mitig,pec_shadow_2023} typically applies Clifford recovery operations
which would be many orders of magnitude less efficient in the present context as we demonstrate below.
In contrast, the present approach uses a large number of operations $\ul$ that are close to the desired operation
-- this however, makes our classical pre-processing step rather challenging as it requires
manipulating ill-conditioned design matrices and thus standard tools from convex optimisation are impractical.
For example, we will consider sequences $\mathcal{U}_l$ of Clifford and T gates  
that approximate the same continuous rotation,
and are thus very close to each other
$\lVert \mathcal{U}_l - \mathcal{U}_k \rVert \leq 10^{-3}$ resulting in an ill-conditioned optimisation problem.

 We address this technical challenge by first posing the decomposition $\gamma_l$ 
as the solution of an underdetermined linear system of equations.
Given underdetermined equations may have infinitely many solutions, our aim is to find the optimal one that has a minimal
$\lVert \gamma \rVert_1$ norm and thus guarantees a minimal measurement overhead of the approach.
In fact, an important result in the theory of linear systems is that the minimal L1-norm
solution is almost always the sparsest~\cite{elad2010sparse,schmidt2007fast}
and this sparsity is advantageous for the following key reasons.
a) the number of different quantum operations used in the experiment does not scale
with the (potentially very large) size of the present classical gate library
and b) the solution can be found iteratively using a number of steps that does
not scale with the size of the (potentially very large) size of the present classical gate library.
 Finding such sparse solutions has become a very well explored and
understood field in mathematics with numerous important applications in engineering, signal processing
(compressed sensing) and in machine learning~\cite{elad2010sparse,schmidt2007fast}.
We employ sophisticated Least Angle Regression (LARS) solvers~\cite{elad2010sparse,loris2008l1packv2} that can achieve arbitrary precision while allowing us to 
efficiently prepare not only the exact solution but an entire, continuous family of solutions 
that make a tradeoff between approximation error and L1 norm (measurement overhead).

We demonstrate a diverse spectrum of practical applications in quantum technologies.
First, we demonstrate the exact synthesis  of continuous-angle rotations  in fault-tolerant quantum computers
 using only short T-count sequences such that the measurement overhead achieved is more than 4 orders of magnitude lower than with previous techniques. 
Second, we demonstrate how the present approach enables exactly or approximately synthesising
broadband and band-selective pulses in quantum optimum control -- which can ultimately lower error rates due to finite-bandwidth
effects in native,  physical quantum gates used in today's quantum computers and quantum simulators.
Third, we extend these results to NMR and MRI applications where broadband and band-selective pulses
are crucial~\cite{levitt2013spin}.
Finally, we considerably expand prior results on overcoming finite rotation-angle resolution due to discretisation
in the classical control electronics.

This manuscript is organised as follows. In \cref{sec:theory} we first review the basic theory
of probabilistic schemes that use quasiprobability decompositions. We then pose the present problem
as solving an underdetermined system of equations, introduce convex optimisation and the numerical solvers one can employ. 
In \cref{sec:appl} we detail a broad range of practical applications with explicit examples.

\section{Theory\label{sec:theory}}

\subsection{Quasiprobability representations}
In the following we refer to a unitary operator $U \in \mathbb{C}^{2^N \times 2^N}$ in terms of its
process matrix $\mathcal{U}$, which acts on a vectorised density matrix isomorphically with conjugation as
$\mathcal{U} \mathrm{vec}[\rho]= \mathrm{vec}[U \rho U^\dagger]$.
To simplify our presentation, we further assume that  $\mathcal{U}$ is the Pauli transfer matrix
such that $\mathcal{U} \in \mathbb{R}^{2^{2N} \times 2^{2N}}$ is a real matrix~\cite{nielsen2010quantum}. 
The exponential in $N$ size of the process matrix
does not pose a significant limitation in practice given the present approach is only
applied to design improved local gate operations, i.e., in most practical applications
quantum gates typically only act on a small subset of all qubits.
Improved general quantum circuits are then obtained as a composition of the improved local,
probabilistic gate operations.

The central aim of this work is to implement a desired, high-cost gate operation $\udes$ 
(e.g., one that the hardware can only implement at a substantial quantum overhead)
as a linear combination of gates $\{\mathcal{U}_l \}_{l}$
that the quantum hardware can natively implement at low cost as
\begin{equation} \label{eq:quasiprob}
	\udes = \sum_{l=1}^{\ndct}   \gamma_l   \mathcal{U}_l.
\end{equation}
For example, we will consider applications where $\udes$ would require an infinitely long
sequence of Clifford and $\mathrm{T}$ gates whereas $\mathcal{U}_l$ are finite-depth approximations.
While the decomposition in \cref{eq:quasiprob} cannot be directly 
realised in an experiment, it can be implemented as the average of the following  random sampling scheme~\cite{koczor2023probabilistic}.
\begin{statement}\label{def:prob_sch}
	We define a sampling scheme that implements $\udes$ on average
	by randomly choosing the gates $\ul$ in our library 
	according to the probability distribution $p(l):= |\gamma_l|/ \lVert \gamma \rVert_1$.
	Then we obtain an unbiased estimator of the desired quantum gate as
	\begin{equation}
		\hat{\mathcal{U}}_{desired} = \lVert \gamma \rVert_1 \mathrm{sign}(\gamma_l) \mathcal{U}_l,
	\end{equation}
	with the property $\mathbb{E}(\hat{\mathcal{U}}_{desired} ) = \mathcal{U}_{desired}$.
\end{statement}
Refer to \cref{app:st1_proof} for a proof.
Similar sampling schemes have been employed in the context of quantum error mitigation
whereby the approach is typically called a quasiprobability representation~\cite{gambetta_error_mitig,practical_QEM,berg2022probabilistic}
due to the possibly negative coefficients $\gamma_l$;
While it is not possible to directly apply a negative weight in a single experiment, one
uses the above scheme to estimate expectation values
whereby individual measurement outcomes are multiplied in post-processing with the sign $\mathrm{sign}(\gamma_l)$ of the relevant circuit variant.
Negative signs then cause an increase in variance and thus an increased number of circuit repetitions
are required for estimating an observable via the L1 norm $\lVert \gamma \rVert_1 = \sum_l |\gamma_l|$.
\begin{statement}\label{statement:var}
Applying \cref{def:prob_sch} to the estimation of the expected value $o = \tr[O \udes \rho ]$ of an observable 
 results in the unbiased estimator $\hat{o}$ such that $\mathbb{E}[\hat{o}] = o$.
The number of shots $N_s$ required to achieve precision $\epsilon$ is upper bounded (assuming norm $\lVert O \rVert_\infty =1$)
as $N_s \leq  \epsilon^{-2} \lVert \gamma \rVert_1^2$.
Building a quantum circuit of $\nu$ probabilistic gate implementations
results in the overhead as
\begin{equation}\label{eq:overhead}
N_s \leq \epsilon^{-2}   \lVert \gamma_{max} \rVert_1^{2\nu},
\end{equation}
where $\lVert \gamma_{max} \rVert_1$ is the largest norm  -- and of course the circuit can additionally contain any number of deterministically chosen gates.
\end{statement}
Refer to \cref{app:var} for a proof that explicitly constructs the estimator $\hat{o}$. 
The above measurement cost is exponentially larger than
having direct access to $\udes$ in which case a constant number of shots $N_s \leq \epsilon^{-2} $
would suffice.
Still, the present approach is highly beneficial in practice
as long as the number $\nu$ of gates in the circuit is not significantly larger than $\lVert \gamma_{max} \rVert_1^{-1}$
-- and indeed the present bounds are quite pessimistic as the actual overhead may be orders of magnitude lower~\cite{koczor2023probabilistic,locality_error_mitigation_ibm_2023}.

Our primary aim is to find decompositions which have an overhead as low as possible -- we achieve this 
via two conceptually different routes. First, we carefully design a library of gates (design matrix)
which by construction provides 	low overhead solutions. Second, out of the infinite family of solutions
we choose the unique one that minimises the L1 norm $\lVert \gamma \rVert_1$ of the solution vector.

\subsection{Constructing a design matrix\label{sec:solve}}
We start by vectorising all process matrices $\mathrm{U}_{desired} = \mathrm{Vec}{[\udes]}$
and $\mathrm{U}_{l} = \mathrm{Vec}{[\ul]}$
by stacking column vectors on top of each other;
We then arrange these (column) vectors $\mathrm{U}_{l}$ into a design matrix as 
\begin{equation} \label{eq:design_mat}
	\textbf{R} = \left( \mathrm{U}_{1}, \mathrm{U}_{2}, \dots \mathrm{U}_{N_{dict}} \right),
	\quad \text{thus} \quad 
	\textbf{R} \underline{\gamma} = \mathrm{U}_{desired},
\end{equation}
where on the right-hand side we re-write \cref{eq:quasiprob} as a matrix-vector equation.
The above is a linear system of equations with an unknown vector $\underline{\gamma}$
and a design matrix of dimension
$\textbf{R} \in \mathbb{R}^{2^{4N} \times \ndct}$.

We detail in the Appendix  that quantum error mitigation~\cite{piveteau2022quasiprobability}
protocols mostly  consider square design matrices $\textbf{R}$ of dimension $2^{4N} = \ndct$
that are well-conditioned and invertible~\cite{practical_QEM,piveteau2022quasiprobability}, e.g.,
a single-qubit gate $N=1$ is mitigated using 16 Clifford operations 
through the unique solution $\underline{\gamma} = \textbf{R}^{-1}  \mathrm{U}_{desired}$.
Unfortunately, mitigating coherent errors using Clifford operations may result in many orders of magnitude larger
overheads than an optimal scheme with minimal overhead: ref~\cite{koczor2023probabilistic} analytically proved
that a minimal L1 norm is achieved when quantum operations $\ul$
are chosen that are as close to the desired operation as possible -- in addition
to possible Clifford operations.

In the present case we aim to find low-overhead solutions by constructing our design matrix of a large number of operations
$\{\mathcal{U}_l \}_{l}$ that are as close to the desired one $\udes$ as possible.
We will thus assume that the column dimension of $\textbf{R}$ is smaller than its row dimension via $2^{4N} \ll \ndct$
and typically $\textbf{R}$ is ill-conditioned.

\subsection{Convex optimisation\label{sec:optimise}}
Given a fixed design matrix as outlined in the previous section,
the system of equations in \cref{eq:design_mat} is underdetermined  and thus may have infinitely many solutions;
our aim is thus to find the unique solution with minimal L1 norm as 
\begin{equation} \label{eq:exact_l1_minimisation}
	\min_{\underline{\gamma}}  \lVert \gamma \rVert_1
	\quad \quad \text{subject to}
	\quad \quad
	\textbf{R} \underline{\gamma} = \mathrm{U}_{desired}.
\end{equation}
Such minimal l1-norm solutions  are a)  almost always the sparsest solutions
and b) can
be solved efficiently using tools from convex optimisation~\cite{elad2010sparse,schmidt2007fast}.
More specifically, we consider the general optimisation problem
that additionally allows us to find approximate solutions as
\begin{equation}\label{eq:approx_opt}
	\min_{\underline{\gamma}} \left(
	\lambda \lVert \gamma \rVert_1
	+
	\lVert \textbf{R} \underline{\gamma} - \mathrm{U}_{desired} \rVert_2
	\right).
\end{equation}
By choosing a sufficiently large $\lambda$ one demands a sparse but approximate solution
while decreasing $\lambda$ yields more accurate solutions at the expense of increasing the norm $\lVert \gamma \rVert_1$.
Solving \cref{eq:approx_opt} has been central to a wide array of practical applications:
the problem is known as basis pursuit denoising (BPDN)~\cite{chen2001atomic}
in signal processing 
while it is known as LASSO in machine learning applications~\cite{simon2013sparse}.
As such, a significant advantage compared to, e.g., generic linear programming solvers is that
fast, specific solvers are available.

\begin{figure}
	\centering
	\includegraphics[width=0.4\textwidth]{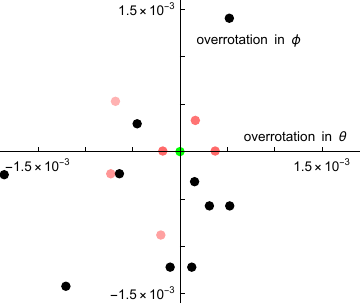}
	\caption{
		Overrotation (black and red dots) in terms of spherical angles $\theta$ and $\phi$  
		of Clifford+T sequences of T count at most $36$
		relative to the targeted continuous angle rotation gate (green dot).
		A design matrix $\mathbf{R}$ was constructed using $20$ Clifford+T sequences
		and a further $15$ Clifford rotations (not shown due to their large overrotations).
		We exactly synthesise the ideal gate (green dot)
		with a minimal measurement cost  $\lVert \gamma \rVert_1 - 1 = 10^{-6.7}$
		that is more than 4 orders of magnitude lower than achievable via prior techniques~\cite{practical_QEM,piveteau2022quasiprobability,suguru_fault_tolerant}.
		This optimal solution is sparse: black dots are rotation gates that 
		receive zero coefficients $\gamma_l = 0$ while red dots are gates with
		non-zero coefficients (their opacity illustrates $|\gamma_l|$).
	}
	\label{fig:cliffordt}
\end{figure}

In the present work we specifically construct design matrices that are ill-conditioned
which may result in numerical precision issues.
For this reason we use a Least Angle Regression (LARS) solver that constructs
very accurate solutions via the following steps~\cite{elad2010sparse,loris2008l1packv2}:
One starts by solving \cref{eq:approx_opt} in the limit $\lambda \rightarrow \infty$
via the null vector $\underline{\gamma} = (0,0, \cdots 0)^T$.
Then the value of $\lambda$ is reduced
such that a single vector
entry is selected that has the highest potential of reducing the error term
$\lVert \textbf{R} \underline{\gamma} - \mathrm{U}_{desired} \rVert$ and a new solution vector is provided 
with only a single non-zero entry. The approach then adds single entries to the vector 
gradually reducing the value of $\lambda$ until an exact solution is found (or $\lambda = 0$ is approached).   
The significant advantage of this approach is that we construct the entire path along $\lambda$.
Thus we can efficiently obtain a solution where the constraint $\lVert \gamma \rVert_1 = 1$ is exactly satisfied,
i.e., no overhead, which is very useful in practice as we demonstrate later.

Furthermore, we note that the error term $\lVert \textbf{R} \underline{\gamma} - \mathrm{U}_{desired} \rVert$ in 
\cref{eq:approx_opt} is the Hilbert-Schmidt distance  between the two process matrices $\textbf{R} \underline{\gamma}$
and $\mathrm{U}_{desired}$.
Recall that the Hilbert-Schmidt distance of two density matrices is an operationally invariant distance
between two fixed quantum states expressing an average distance in terms of measurement probabilities~\cite{lee2003operationally}.
Similarly, the Hilbert-Schmidt distance between two process matrices expresses the average distance between two
quantum processes as detailed in~\cite{maciejewski2023exploring}.

\section{Applications\label{sec:appl}}
\subsection{Application 1: Fault-tolerant gate synthesis\label{sec:clifft}}
Fault-tolerant quantum computers typically achieve universality through
decomposing continuous rotation gates into sequences of 
(relatively) expensive T-gates and (relatively) cheaper Clifford gates; well-established algorithms are available that efficiently 
decompose any single-qubit rotation into such sequences~\cite{ross2014optimal}. 
However, in early fault-tolerant quantum computers one will make frugal use of T gates
-- the limited T count of Clifford+T sequences thus introduces a coherent deviation
from the desired quantum operation.

We build on the observation that a large number of
Clifford+T sequences can be constructed that contain at most $N_T$ T gates
and approximate the ideal gate $\udes$ to a precision at least $\epsilon$.
For example,  we consider a single-qubit $Z$ rotation $R_z(0.234234)$
and explicitly construct a library of $20$ different Clifford+T sequences as $\ul$ 
that achieve a precision (Hilbert-Schmidt distance) at best $10^{-3.4}$ using at most $36$ T gates.
The design matrix is then obtained by appending a further 15 single-qubit Clifford
operations to one of the $\ul$ such that $\mathbf{R}  \in \mathbb{R}^{35 \times 16}$.
\cref{fig:cliffordt} (black and red dots)  illustrates the overrotation in spherical angles $\theta$ and $\phi$
with respect to the ideal operation (green) when applying elements of the gate library onto an initial quantum state.

By solving the optimisation problem in \cref{eq:exact_l1_minimisation}
for $\lambda \rightarrow 0$ we obtain an exact solution that allows to exactly synthesise the desired $R_z$ rotation gate (green dot) at a minimal
measurement overhead $\lVert \gamma \rVert_1 - 1 = 10^{-6.7}$
 -- this overhead is more than 4 orders of magnitude lower than 
using Clifford recovery operations only as in \cite{practical_QEM,piveteau2022quasiprobability,suguru_fault_tolerant} 
in which case $\lVert \gamma \rVert_1 - 1 = 10^{-2.3}$.
Furthermore, \cref{fig:cliffordt}  illustrates that indeed our solution is sparse as most gates in the library receive
zero coefficients (black dots, $\gamma_l =0$) while  red dots correspond to elements of the library that receive non-zero 
coefficients while their opacity is proportional to the absolute values of $\gamma_l$.

We note that in the present example we used the simple and optimal approach of ref~\cite{ross2014optimal}
that requires no ancilla qubits and requires $\approx 3.02\log_{2} (1/\epsilon )+1.77$ T gates to
achieve precision $\epsilon$. One can reduce the number of T gates
to $\approx 1.03\log_{2} (1/\epsilon )+5.75$
via the probabilistic approach of \cite{PhysRevA.91.052317} that requires ancilla qubits,
resulting in $17$ T gates used in the present example rather than $36$.

\begin{figure}
	\centering
	\includegraphics[width=0.4\textwidth]{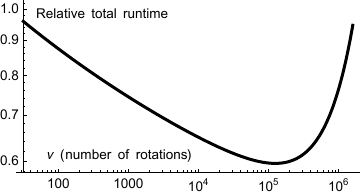}
	\caption{
		Estimated relative total runtime, as a ratio of relative total T counts, 
			with respect 
			to direct synthesis for an increasing number $\nu$
			of rotation gates.
			Even when we take into account its measurement overhead, the present
			approach requires in total fewer T gates than direct synthesis
			as long as the number $\nu$ of gates is less than $(\lVert \gamma \rVert_1 - 1)^{-1}$.
			Furthermore, the present approach 			
			has the significant advantage that it guarantees zero approximation
			error, whereas direct synthesis is approximate and we considered
			a very permissive circuit-level error $\epsilon_{circ} = 10^{-2}$.		
	}
	\label{fig:runtime}
\end{figure}

	Let us compare the end-to-end time complexity of the present
	approach with direct synthesis assuming the technique of ref.~\cite{PhysRevA.91.052317}.
	Assuming $\nu$ continuous rotation gates yields a multiplicative
	increase of $\exp{(4.0\times 10^{-7}\times \nu)}$ in the number of circuit repetitions via \cref{eq:overhead}.
	Given our library consists of rotations of at best $\epsilon_p=10^{-3.4}$,
	we need $1.03\log_{2} (1/\epsilon_p)+5.75 \approx 17$ 
	T gates per rotation which results in a T count of $17 \nu $ of a single circuit execution.
	Then, considering the measurement overhead we obtain a total T count of
	\begin{equation}
		17 \nu \exp{(4.0\times 10^{-7}\times \nu)}.
	\end{equation}

	In contrast, using direct synthesis with a realistic $\epsilon_d = \nu^{-1} 10^{-2}$ requires 
	$1.03\log_{2} (1/\epsilon_d)+5.75 $
	T gates per rotation resulting in a total T count of 
	\begin{equation}
		12.6 \nu + 1.03 \nu \log_{2} (\nu).
	\end{equation} 
	We plot the relative total T counts as an estimate of the relative runtime of both approaches
	for an increasing number $\nu$ of gates in \cref{fig:runtime}.
	Indeed, as long as the number $\nu$ of continuous rotation gates in the circuit is smaller than 
	$(\lVert \gamma \rVert_1 - 1)^{-1}$, 
	the present approach achieves a lower total T count which can therefore reduce the runtime of quantum algorithms.
	However, a significant further advantage of the present approach is that it guarantees zero approximation
	error $\epsilon=0$,
	in contrast to direct synthesis for which we assume a permissive circuit level error
	$\epsilon_{circ} = 10^{-2}$ in the present example.
	While the example in \cref{fig:runtime} is limited to circuits with no more than $\approx 10^7$ rotation gates,
	we can straightforwardly extend the approach 
	by building a new library of rotations using a smaller $\epsilon_p$ (given in the present example we
	used $\epsilon_p=10^{-3.4}$
	which resulted in $\lVert \gamma \rVert_1 - 1 = 10^{-6.7}$).

\begin{figure*}
	\centering
	\includegraphics[width=\textwidth]{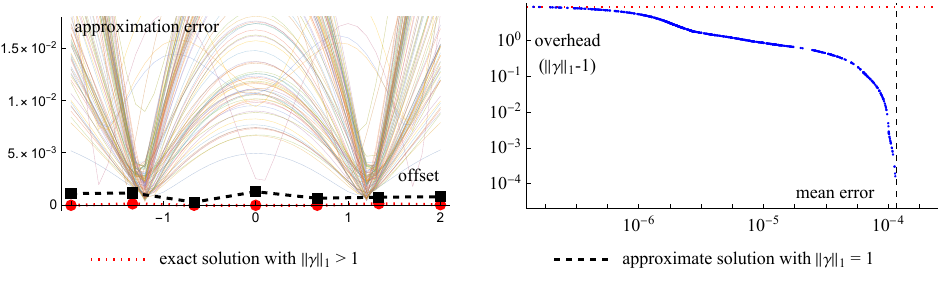}
	\caption{
		A library of 100 different optimised pulses was constructed
		that all approximate an ideal $\udes = \mathcal{R}_x(\pi/2)$ rotation.
		(left, solid lines)
		approximation error  $\lVert \ul(d) - \udes \rVert_2$  of each optimised pulse
		 as a function of the drift (offset) $d$ in \cref{eq:ham}.
		We obtain sparse solutions that implement exactly (red) and approximately (black) the desired rotation for all considered drift values (black squares and red dots).
		The approximate solution (black dashed) significantly outperforms any of the optimised pulses (solid lines)
		yet it does not require an increased number of samples via  $\lVert \gamma \rVert_1 = 1$
		-- in contrast, the exact solution introduces a measurement overhead $\lVert \gamma \rVert_1 \approx 8.65$. 
		(right) tradeoff curve showing the measurement overhead  $\lVert \gamma \rVert_1 - 1$
		for solutions of decreasing sparsity
		as a function of the approximation error averaged over the different drift values.
	}
	\label{fig:opt_cont}
\end{figure*}

\subsection{Application 2: Quantum optimum control}
Physical quantum gates are typically implemented in hardware
by electronically controlling the interaction strength between  
qubits and external electromagnetic fields.
Sophisticated optimum control techniques have been developed for 
numerically discovering non-trivial control-pulse shapes~\cite{werschnik2007quantum,koch2022quantum}.
Given each instance of the numerical search typically finds a local optimum,
we can build a library of these different optimum control solutions $\ul$;
By solving  \cref{eq:approx_opt} we can then either exactly, or approximately but with lower approximation error
than any of the $\ul$,
implement the desired gate on average.

To simplify our presentation, let us consider the specific
problem of synthesising a single-qubit rotation $\udes = \mathcal{R}_x(\pi/2)$---while of course the approach applies generally to any other optimum control
scenario---via the time-dependent Hamiltonian as
\begin{equation}\label{eq:ham}
	\mathcal{H}(t_k, d) = d H_0 + \re [h(t_k)] X + \im[ h(t_k)] Y.
\end{equation}
Here the piecewise-constant, complex parameters $h(t_k)$ 
control the strength of the Pauli $X$ and $Y$ interactions for
each timestep $t_k$ and typically
correspond to the phase and amplitude of an applied RF or MW pulse.
Furthermore, $H_0$ is a drift Hamiltonian, i.e., an interaction that the hardware cannot directly control.
Implementing the piecewise constant pulse $\mathcal{H}(t_k,d)$ in which 
each piece has length $\Delta t$  yields the unitary gate as 
\begin{equation*}
	U(d) = \prod_{k=1}^l e^{-i \mathcal{H}(t_k,d)  \Delta t }.
\end{equation*}
Optimum control techniques numerically  optimise the parameters
$h(t_k)$ in order to maximise the fidelity of the emerging unitary gate $F(d) = | \tr [  R_x(\pi/2) U^\dagger(d) ]|$ 
that depends on the drift $d$  -- broadband pulses then maximise $F(d)$ for a broad range of $d$ values.

\cref{fig:opt_cont} (left, solid lines) shows approximation errors (Hilbert-Schmidt distance) 
in a library of $N_{dict} = 100$ shaped pulses that were optimised to achieve maximal fidelity
in the range of dimensionless offset values $-2 \leq d \leq 2$.
Exactly achieving an $R_x(\pi/2)$ rotation for a broad range of offsets requires arbitrarily large amplitudes $h(t_k)$,
however, in order to limit the energy dissipation of the quantum hardware it is highly desirable to 
cap the maximum allowed amplitude.
The optimised pulses in \cref{fig:opt_cont} (left, solid lines)
were capped at $|h| \leq 6$ and thus the approximation error is non-zero but minimal.
Indeed, as each pulse optimisation was initialised in a different random
seed, each converged to a different local optimum resulting in a different process matrix $\ul(d)$ and a different error-distribution curve in \cref{fig:opt_cont} (left, solid lines).
We build a library of these process matrices and apply our probabilistic approach to exactly implement on average the desired rotation
for a range of offset values $-2 \leq d \leq 2$.

For this reason, we first discretise the offset range into $q$ discrete values $(d_1, d_2, \dots d_q)$
and stack the corresponding vectorised process matrices on top of each other as
\begin{equation*}
	 \mathrm{U}_{l} = 
	 \begin{pmatrix}
	 	\mathrm{Vec}{[\ul(d_1)]}\\
	 	\mathrm{Vec}{[\ul(d_2)]}\\
	 	\vdots\\
	 	\mathrm{Vec}{[\ul(d_q)]}
	 \end{pmatrix},
	 \quad	\mathrm{U}_{desired} =
	 \begin{pmatrix}
	 	\mathrm{Vec}{[\udes]}	\\
	 	\mathrm{Vec}{[\udes]}	\\
	 	\vdots\\
	 	\mathrm{Vec}{[\udes]}	
	 \end{pmatrix}.
\end{equation*}
The vectors $\mathrm{U}_{l}$ on the left-hand side then form columns of our design matrix
$\textbf{R} = \left( \mathrm{U}_{1}, \mathrm{U}_{2}, \dots \mathrm{U}_{N_{dict}} \right)$
as in \cref{eq:design_mat}.
Thus, discretising the offset range into $q$ pieces results in a column dimension
of the design matrix $q 2^{4n}$ that has been increased by a factor of $q$.
The vector of the desired process $\mathrm{U}_{desired}$ (right-hand side of the above equation)
then simply consists of  $q$ copies of the vectorised process matrix of the desired operation.
The linear system of equations $\mathrm{U}_{desired} = \textbf{R} \underline{\gamma}$
must have thus solutions $\underline{\gamma}$
that guarantee that the same $\mathrm{Vec}{[\udes]}$ is obtained for all offset values $(d_1, d_2, \dots d_q)$.

\cref{fig:opt_cont}(left, red dots) shows the exact solution for $\lambda \rightarrow 0$
using the design matrix of size $\textbf{R}  \in \mathds{R}^{112 \times 340}$
that was constructed by discretising the offset range into $q=7$ grid points
for $100$  different shaped pulses $\ul$ and appending additional 
Clifford operations (achieved via, e.g.,  phase shifts of the pulses).
As detailed in \cref{sec:optimise}, the $\lambda \rightarrow 0$ solution is constructed by
starting at $\lambda \rightarrow \infty$ and gradually decreasing
the sparsity of the vector $\vec{\gamma}$ which monotonically decreases the error while also monotonically
increasing the  norm $\lVert \gamma \rVert_1$. The resulting trade-off curve is shown in
\cref{fig:opt_cont}(right) where the approximation error (x axis) is plotted against the L1 norm (y axis)
-- the L1 norm of the aforementioned exact solution is shown by the dotted red line.

In applications where a lower measurement cost is desired, an approximate solution of small L1 norm may be preferred.
Among these, one can obtain a solution that achieves zero measurement overhead via $\lVert \gamma \rVert_1 =1$, i.e.,
\cref{fig:opt_cont}(right, dashed black).
The average approximation error over the relevant drift values is illustrated in
\cref{fig:opt_cont}(left, dashed black) which confirms that indeed the approximate solution outperforms
any of the individual optimised pulses $\ul(d)$ (lower approximation error than solid lines)
even though no measurement overhead is introduced.

A number of natural generalisations and further applications are apparent. For example, 
following the same steps one can construct band-selective pulses
via the vectorised process matrix $U_{desired}$ as
\begin{equation*}
	\mathrm{U}_{desired} =
	\begin{pmatrix}
		\mathrm{Vec}{[\openone]}	\\
		\mathrm{Vec}{[\openone]}	\\
		\vdots \\
		\mathrm{Vec}{[\udes]}	\\
		\mathrm{Vec}{[\udes]}	\\
		\vdots\\
		\mathrm{Vec}{[\udes]}	\\
		\vdots\\
		\mathrm{Vec}{[\openone]}	\\
		\mathrm{Vec}{[\openone]}
	\end{pmatrix}.
\end{equation*}
This construction of $\mathrm{U}_{desired} $ can be used guarantee that $\udes$ is applied only within the
offset range $-B \leq d \leq B$ while
the identity operation is applied otherwise.

\subsection{Application 3: broadband and band-selective pulses in NMR and MRI}
The techniques introduced in the previous section naturally extend to NMR and MRI applications,
where broadband and band-selective pulses are essential.
We illustrate our approach on the specific application where an ideal, $\pi/2$ rotation
$\udes = \mathcal{R}_x(\pi/2)$ needs to be implemented (which is again represented via its Pauli transfer matrix);
This gate operation is used to transform the average state  of a very large number of nuclear 
or electron spins as the mixed state $\rho$ as~\cite{levitt2013spin}
\begin{equation} \label{eq:state_transf}
	\ul \mathrm{vec}[\rho] = \mathrm{vec}[\rho'_l].
\end{equation}
By convention, the initial state $\rho \propto Z$ is proportional to the Pauli $Z$ matrix
(ignoring the identity term in $\rho$ that that does not contribute to the
measurement outcome~\cite{levitt2013spin}).
Optimal control techniques can be used to obtain pulses $\ul$ that best approximate the ideal action 
$\udes = \mathcal{R}_x(\pi/2)$ in order to obtain the transformed state  $\rho'_{desired} \propto Y$
that is proportional to the Pauli $Y$ matrix.

A simple NMR or MRI experiment then proceeds by allowing this state to freely evolve for time $t$ into $\rho_l'(t)$ under the natural Hamiltonian
of the spin system; Measuring Pauli $X$ and $Y$ operators yield real and imaginary parts of the classical time-dependent
signal $S_l(t)$ as
\begin{equation}\label{eq:signals}
	S_l(t) = \tr[ \rho_l'(t) X ] + i \tr[ \rho_l'(t) Y].
\end{equation}
Finally, the experiment may be repeated $N_s$ times in order to suppress statistical noise in $S_l(t)$ by a factor $N_s^{-1/2}$
and the resulting average signal is Fourier transformed to obtain an NMR spectrum.
However, as explained in the previous section, the pulses $\ul$ illustrated in \cref{fig:opt_cont} (solid lines)
incur an approximation error due the drift term $d \mathcal{H}_0$
and thus the state $\rho_l'$ is not exactly the desired one $\rho'_{desired} \propto Y$.

We can adapt our probabilistic scheme in \cref{def:prob_sch} such that
we randomly choose pulses $\ul$ to be implemented in an NMR or MRI experiment.
\begin{statement}\label{stat:nmr}
Given an estimator of a desired operation $\udes$ from \cref{def:prob_sch},
we choose operations $\ul$ according to the
probability distribution  $p(l):= |\gamma_l|/ \lVert \gamma \rVert_1$
and perform an NMR/MRI experiment. An unbiased estimator of the ideal NMR/MRI signal 
is obtained as
\begin{equation}
\hat{S}(t)_{desired} = \lVert \gamma \rVert_1 \sign(\gamma_l) S_l(t),
\end{equation}
in terms of the individually measured signals $S_l(t)$ from \cref{eq:signals}
 such that $\mathbb{E}[\hat{S}(t)_{desired}] = S(t)_{desired}$.
\end{statement}
Refer to \cref{proof_nmr} for a proof.
The above statement results in the following practical protocol:
We first randomly choose a pulse $\ul$ and use it to prepare the initial state via \cref{eq:state_transf},
we measure the signal $S_l(t)$ as a function of the evolution time as in~\cref{eq:signals}, 
multiply each signal with the relevant sign and the L1 norm of the solution vector;
The mean of the resulting signals is then guaranteed to recover the ideal signal $S_{desired}(t) = \tr[\rho'_{desired} (X +i Y)]$
that one would obtain by applying the ideal gate at each offset value $d$, i.e., red dots and black squares in \cref{fig:opt_cont}.

As an NMR experiment is repeated $N_s$ times to suppress random noise by a factor $N_s^{-1/2}$,
indeed the number of repetitions required to achieve precision $\epsilon$ scales as $N_s \propto \epsilon^{-2} \lVert \gamma \rVert_1^2$
just like in the case of quantum computing applications in \cref{statement:var}.
Furthermore, complex NMR pulse sequences may apply multiple different gate elements one after the other -- one would then
replace each gate element with the relevant estimators from \cref{stat:nmr}, the same way as quantum circuits are constructed in \cref{statement:var};
The measurement overhead then similarly grows exponentially with the number of gate elements.

\subsection{Application 4: optimality of Probabilistic Angle Interpolation}

Most leading quantum computing platforms require cryogenic cooling which introduces a communication bottleneck --
hardware developers thus aim to move control electronics closer to the QPU which, however,
introduces limitations  to the classical control electronics, such as 
low-precision discretisation of continuous rotation gates~\cite{koczor2023probabilistic}.
PAI resolves this issue~\cite{koczor2023probabilistic} 
by effectively implementing a continuous-angle rotation gate $\mathcal{R}(\cdot)$  
when the hardware platform can only implement a discrete set of
rotations $\ul = \mathcal{R}(\Theta_l)$ via the angles as $\Theta_l = l 2 \pi/ 2^B$ with $l \in \{ 0, \dots 2^B -1 \}$.

We use the present approach to verify that the analytical solution of ref~\cite{koczor2023probabilistic}
solves the exact optimisation problem in \cref{eq:exact_l1_minimisation}
and also to obtain further, approximate solutions via the generalised optimisation problem in \cref{eq:approx_opt}.
In particular, ref~\cite{koczor2023probabilistic} obtained an analytical decomposition 
whereby only 3 gate configurations receive non-zero coefficients $\gamma_l$, i.e.,
one randomly chooses one of the two notch settings $\Theta_k$ and $\Theta_{k+1}$ nearest
to the desired angle $\theta$ as well as a third setting that corresponds to the polar opposite rotation $\Theta_k + \pi$.

\begin{figure}
	\centering
	\includegraphics[width=0.35\textwidth]{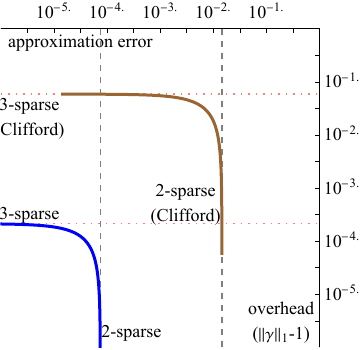}
	\caption{
		PAI implements an arbitrary rotation gate $\mathcal{R}(\theta)$  
		when only $2^B$ discrete rotation angles are accessible $\ul = \mathcal{R}(\Theta_l)$ (we assume $B=7$ here).
		An exact, analytical solution  was found in ref~\cite{koczor2023probabilistic}
		that is 3-sparse with overhead $\lVert \gamma \rVert_1-1 = 10^{-3.67}$ (blue solid, red dotted).
		We additionally find a 2-sparse solution with no measurement overhead
		but with an approximation error $\lVert \textbf{R} \underline{\gamma} - \mathrm{U}_{desired} \rVert_2  = 10^{-4.12}$ (blue solid, grey dashed). 
		(blue solid line) Interpolating between these two solutions allows us to obtain an infinite family of solutions
		that offer a tradeoff between low approximation error and low measurement overhead.
		(brown solid line) using Clifford recovery operations only as in ref~\cite{suguru_fault_tolerant}
		results in orders of magnitude higher measurement cost (and/or approximation error).
	}
	\label{fig:pai}
\end{figure}

We illustrate our approach by considering a single-qubit Pauli X rotation at discretised rotation angles $\ul = \mathcal{R}(\Theta_l)$
with $B=7$ bits of precision and build a design matrix of size $\textbf{R} \in \mathbb{R}^{128 \times 16}$
that contains all $128$ discrete gate variants; we find the following conclusions.
First, we find that a 3-sparse solution exactly implements the desired gate ($\lambda \rightarrow 0$)
and confirm that this solution is identical to the analytical solution of ref.~\cite{koczor2023probabilistic}.
Second, the significant advantage of the present approach is that we can additionally obtain approximate solutions in
 \cref{eq:approx_opt} via $0 < \lambda < \infty$
that have lower measurement cost.
As such, we obtain a  2-sparse solution that achieves zero measurement overhead as $\lVert \gamma \rVert_1 = 1$.
Interestingly, this (optimal) solution achieves slightly lower approximation error than a naive solution whereby one chooses 
$\Theta_k$ with probability $p$ and chooses $\Theta_{k+1}$ with probability $1-p$, where
$p$ is the relative position of the desired angle $\theta$ to the two nearest notch settings.

Third, by linear combining the 2-sparse and 3-sparse solution vectors, we obtain an infinite family
of solutions that make a tradeoff between approximation error and measurement overhead (\cref{fig:pai}, blue solid).
In comparison, using only Clifford recovery operations as in \cite{suguru_fault_tolerant} results in orders
of magnitude higher measurement overheads and approximation errors (\cref{fig:pai}, blue solid).

\section{Discussion and conclusion}
The present work develops an approach that allows one to synthesise quantum operations
that would otherwise be impossible to implement directly in quantum hardware.
This is achieved by probabilistically interpolating between nearby quantum operations that can be 
realised natively by the quantum hardware.

Probabilistic schemes have been used in quantum error mitigation for mitigating gate errors and have also been used
to overcome coherent errors in unitary gates.
However, previous approaches used Clifford recovery operations 
which allow for finding decompositions via well-conditioned design matrices.
Presently, however, we consider a library of 
gate operations that are close to each other, e.g., gates or gate sequences that approximate the ideal operation.
This guarantees a minimal measurement overhead of the approach, however,  
makes the numerical pre-processing step challenging as one needs to solve a non-trivial underdetermined system of equations.
By using sophisticated tools from convex optimisation we obtain numerically exact, arbitrary-precision solutions.
Furthermore, as we demonstrate, the present approach achieves orders of magnitude lower measurement
overheads than using Clifford-based techniques in the context of quantum computing.
A significant advantage of the present approach is that it is directly compatible with both error mitigation and with 
advanced, randomised measurement techniques,
such as classical shadows~\cite{shadow_huang_2020}
(via the approach of~\cite{pec_shadow_2023}) and will thus benefit a broad set of applications~\cite{shadow_spec_2023,covar}.

Furthermore, the present approach is very general and can be applied to
a broad range of application scenarios beyond quantum computing: we explicitly  demonstrate practical applications
as synthesising perfect (or approximate) broadband and band-selective pulses in quantum optimum control --
as we explicitly demonstrate, these results naturally extend to NMR and MRI applications.

The present work is a promising new direction in developing advanced quantum protocols whereby
quantum resources are significantly reduced through the use of classical computers
that solve non-trivial but offline, classical pre- and post-processing tasks.
While related probabilistic techniques are routinely used in the context of quantum error mitigation, 
the present work extends these mathematical concepts to a spectrum of new application
scenarios in quantum technologies. 

\noindent \textbf{Data availability:} An example code is available online at \url{https://github.com/BalintKoczor/probsynth}.

\section*{Acknowledgments}
BK thanks UKRI for the Future Leaders Fellowship
project titled Theory to Enable Practical Quantum Advantage (MR/Y015843/1).
BK thanks the University of Oxford for
a Glasstone Research Fellowship and Lady Margaret Hall, Oxford for a Research Fellowship.
The numerical modelling involved in this study made
use of the Quantum Exact Simulation Toolkit (QuEST), and the recent development
QuESTlink~\cite{QuESTlink} which permits the user to use Mathematica as the
integrated front end, and pyQuEST~\cite{pyquest} which allows access to QuEST from Python.
We are grateful to those who have contributed
to all of these valuable tools. 
The authors would like to acknowledge the use of the University of Oxford Advanced Research Computing (ARC)
facility~\cite{oxford_arc} in carrying out this work
and specifically the facilities made available from the EPSRC QCS Hub grant (agreement No. EP/T001062/1).
The author also acknowledges funding from the
EPSRC projects Robust and Reliable Quantum Computing (RoaRQ, EP/W032635/1)
and Software Enabling Early Quantum Advantage (SEEQA, EP/Y004655/1).

\appendix
\section{Related works}
\subsection{Quantum error mitigation}
Reducing the sampling overhead has been explored in the context of quantum error mitigation (QEM)~\cite{piveteau2022quasiprobability}:
the diamond distance $\lVert \mathcal{R}_{desired} - \sum_{l}   \gamma_l   \mathcal{U}_l \rVert_\diamond$
was minimised under the constraint that the overhead $\lVert \gamma_l \rVert_1 \leq \epsilon$ stay below a threshold
--
this problem can be solved efficiently using semidefinite programming.
Furthermore, as explained in the main text in QEM one typically chooses the design matrix as a well-conditioned, invertible square matrix, e.g., for a single
qubit one uses 16 Clifford operations~\cite{practical_QEM,piveteau2022quasiprobability}, such that
\cref{eq:design_mat} can be solved straightforwardly as $\underline{\gamma} = \textbf{R}^{-1}  \mathrm{U}_{desired}$.

In the present case we use a large number of gate operations that are very close to each other resulting in a highly ill-conditioned design
matrix whose row dimension is larger than its column dimension, i.e., a non-square matrix.
For this reason we use the Hilbert-Schmidt distance to quantify discrepancy between operations,
rather than the diamond distance of~\cite{piveteau2022quasiprobability}, which allows us to pose 
the present problem as solving an underdetermined system of equations.
Posing the problem this way is a key enabler as we can use precise, efficient special-purpose solvers rather than generic SDP solvers
-- achieving high numerical precision with generic SDP solvers may be difficult.

For this reason we consider applications where one
assumes perfect unitary operations and the only source of error
is due to, e.g., finite T count or discretisation of quantum rotation angles etc.
Furthermore, operations $\mathcal{R}_{desired}$ and $\mathcal{U}_l$ are
known exactly, hence arbitrary precision numerical representations can be used.
This is all in contrast to QEM~\cite{piveteau2022quasiprobability}
whereby one approximately learns these from experiments 
and one thus needs to consider low-precision numerical representations.

While in the main text we assumed the gates are noise free, indeed, in a realistic experiment
the gate elements $\ul$ are likely not perfect. A natural error mitigation strategy was outlined in \cite{koczor2023probabilistic}
that immediately applies to the present approach. In particular, one can efficiently learn the noise model
for each gate variant $\tilde{\ul}$ and probabilistically implement noise-free variants
using Probabilistic Error Cancellation. Hence, the present approach remains unchanged, except, its
measurement cost $\lVert \gamma \rVert_1 $ is increased by a factor due to PEC.

\subsection{Probabilistic gate synthesis}

Probabilistic gate sequences have been employed to improve the accuracy of Clifford+T sequences~\cite{PhysRevA.95.042306,kliuchnikov2022shorter,akibue2023probabilistic}.
In particular, the desired unitary operation is approximated as
$\udes \approx \sum_{l}   p_l   \mathcal{U}_l$ where $p_l$ is a probability distribution
and $\mathcal{U}_l$ are is a set of operations close to $\udes$ but with finite T count.
It was observed that the approximation error in $\sum_{l}   p_l   \mathcal{U}_l$ is always
quadratically smaller than in $\mathcal{U}_l$.
However, these techniques are approximate and as one increases the number of gates
in the circuit, the fidelity of the quantum state decreases exponentially as demonstrated in \cite{koczor2023probabilistic}.

In contrast, the present approach
allows to \emph{exactly synthesise} $\udes$ through the use of both positive and negative weights $\gamma_l$
in \cref{eq:quasiprob};
when the aim is to estimate expected values the $-1$ sign can be applied in post processing.
Furthermore, in contrast to those prior techniques, in our case $\mathcal{U}_l$ additionally include polar opposite rotations
given these guarantee the minimal measurement overhead~\cite{koczor2023probabilistic}.
Indeed, as our approach exactly implements the desired gate, it comes with
an increased measurement overhead that depends quadratically on the
precision, i.e.,   $\min_l \lVert \mathcal{U}_l - \udes \rVert$.

\subsection{Cooperative optimum control}
It was originally proposed in ref~\cite{braun2010cooperative} that shaped pulses $\ul$ that implement a desired rotation
need not be individually accurate but rather need only satisfy a relaxed condition that the average of a series of pulse variants
are required to be accurate over many repeated of measurement rounds.
The present approach is indeed quite comparable, however, we allow for the additional freedom
that different gate/circuit variants are not simply averaged over but we rather
calculate a weighted average where we allow for negative weights even.

\section{Proofs}
\subsection{Proof of \cref{def:prob_sch}\label{app:st1_proof}}
\begin{proof}
We can explicitly evaluate the expected value as
\begin{equation*}
	\mathbb{E}(\hat{\mathcal{U}}_{desired} ) = 
	\sum_l p(l) \hat{\mathcal{U}}_{desired}
	= \sum_l p(l)  \lVert \gamma \rVert_1 \sign(\gamma_l) \mathcal{U}_l,
\end{equation*}
and upon substituting $p(l):= |\gamma_l|/ \lVert \gamma \rVert_1$
we obtain
\begin{equation*}
	\mathbb{E}(\hat{\mathcal{U}}_{desired} ) 
	= \sum_l \gamma_l \sign(\gamma_l) \mathcal{U}_l
	= \mathcal{U}_{desired}
\end{equation*}
where in the last equation  we used \cref{eq:quasiprob}.
\end{proof}

\subsection{Proof of \cref{statement:var} \label{app:var}}
\subsubsection{Unbiasedness}
\begin{proof}
We first start by introducing the measurement  of an observable $O = \sum_b E_b \tr[O E_b]$ via a set of POVMs $\{E_b\}_b$
admitting the property $\sum_b E_b = \openone$. The expected value of an observable is
obtained via the estimator $\hat{x} = \tr[O E_b] $ via the corresponding probability distribution $q_b = \tr[\rho E_b]$.

For the present approach, we construct the estimator $\hat{o}$ via \cref{def:prob_sch} as
\begin{equation*}
	\hat{o} = \lVert \gamma \rVert_1 \sign(\gamma_l)  \tr[E_b O],
\end{equation*}
and verify that it is unbiased as
\begin{equation}
	\mathbb{E}[\hat{o}] = \sum_{l,b} p(l) q_b \hat{o} =\sum_{l,b} p(l) q_b \lVert \gamma \rVert_1 \sign(\gamma_l)  \tr[E_b O].
\end{equation}
Substituting the probabilities $p(l):= |\gamma_l|/ \lVert \gamma \rVert_1$ and  $q_b = \tr[ E_b \ul(\rho)]$ then yields
\begin{align}
	\mathbb{E}[\hat{o}] &= \sum_{l,b}  \gamma_l \tr[ E_b \ul(\rho)]  \tr[E_b O]\\
	&=  \sum_{l}  \gamma_l \tr[ \left( \sum_b \tr[E_b O] E_b\right) \ul(\rho)] .
\end{align}
Here we can substitute $\sum_b \tr[E_b O] E_b = O$ and thus obtain
\begin{equation}
	\mathbb{E}[\hat{o}]  =  \sum_{l}  \gamma_l \tr[ O \ul(\rho)] = \tr[ O \udes(\rho)],
\end{equation}
where the last equation follows from \cref{eq:quasiprob} via linearity of the trace.
\end{proof}

\subsubsection{Bounding the number of shots}
\begin{proof}
For an unbiased estimator, the variance can be upper bounded as
\begin{equation*}
	\var[\hat{o}] = \mathbb{E}(\hat{o}^2) - \mathbb{E}(\hat{o})^2 \leq \mathbb{E}(\hat{o}^2)
\end{equation*}
We thus need only bound the expected value of the square of the estimator as
\begin{align*}
	\mathbb{E}(\hat{o}^2) 
	=& \sum_{l,b} p(l) q_b \hat{o}^2 =\sum_{l,b} p(l) q_b \lVert \gamma \rVert_1^2   \tr[E_b O]^2\\
	\leq& \sum_{l,b} p(l) q_b \lVert \gamma \rVert_1^2   \lVert O \rVert_\infty^2 
	= \lVert \gamma \rVert_1^2   \lVert O \rVert_\infty^2 
\end{align*}
where in the second equation we used the inequality $\tr[E_b O]^2 \leq \lVert O \rVert_\infty^2$.
Finally, the bound on the variance follows as
\begin{equation*}
	\var[\hat{o}] \leq  \lVert \gamma \rVert_1^2   \lVert O \rVert_\infty^2.
\end{equation*}
Thus the number of samples $N_s$ required to achieve a precision $\epsilon$ scales as
\begin{equation}
	N_s = \epsilon^{-2} \var[\hat{o}]  \leq \epsilon^{-2} \lVert O \rVert_\infty^2 \lVert \gamma \rVert_1^2 
\end{equation}

Adapting the proof technique of Statement 3 in \cite{koczor2023probabilistic},
it is straightforward to extend this bound to the case of quantum circuits composed of $\nu$ 
applications of a quasiprobability decomposition in which case the variance is upper bounded
as 
\begin{equation}
	\var[\hat{o}_\nu] \leq   \lVert O \rVert_\infty^{2} \prod_{k=1}^\nu \lVert \gamma_k \rVert_1^{2},
\end{equation}
where $\lVert \gamma_k \rVert_1^{2}$ correspond to the individual quasiprobability decompositions
and is thus bounded by the largest one $\lVert \gamma_{max} \rVert_1$, therefore
$\var[\hat{o}_\nu] \leq \lVert O \rVert_\infty^{2}    \lVert \gamma_{max} \rVert_1^{2\nu}$.
Thus, the number of samples scales in the worst case as
\begin{equation}
	N_s = \epsilon^{-2} \var[\hat{o}_\nu]  \leq \epsilon^{-2}  \lVert O \rVert_\infty^2 \lVert \gamma_{max} \rVert_1^{2\nu}
\end{equation}

\end{proof}

\subsection{Proof of \cref{stat:nmr}\label{proof_nmr}}
We can prove unbiasednes of the estimator by computing the expected value
\begin{equation}
	\mathbb{E}[\hat{S}(t)_{desired}] = 
	\sum_l  p(l) \lVert \gamma \rVert_1 \sign(\gamma_l) S_l(t)
\end{equation}
Substituting the probability $p(l):= |\gamma_l|/ \lVert \gamma \rVert_1$
results in 
\begin{align}\label{eq:1}
	\mathbb{E}[\hat{S}(t)_{desired}]  &= 
	\sum_l  |\gamma_l|  \sign(\gamma_l) S_l(t)\\
	&= \nonumber
	\sum_l  \gamma_l \Big(   \tr[ \rho_l'(t) X ] + i \tr[ \rho_l'(t) Y] \Big)
\end{align}
where in the second equation we substituted the definition of $S_l(t)$
from \cref{eq:signals}. We can simplify the right-hand side by denoting $X$ or $Y$ as $O$ as

\begin{align}\label{eq:2}
	\sum_l  \gamma_l    \tr[ \rho_l'(t) O ] &=
	\tr[ \Big( \sum_l  \gamma_l   \rho_l'(t) \Big)  O ] \\
	&= \nonumber
	\tr[ \Big( \sum_l  \gamma_l   [\ul \rho](t) \Big)  O] \\
	&= \nonumber
	\tr[    [\udes (\rho)](t) \,   O ]
\end{align}
where in the second equation we 
exploited that $\sum_l  \gamma_l \ul = \udes$ and then used the definition of $\rho_l$ from \cref{eq:state_transf}.
By combining \cref{eq:1} and \cref{eq:2} it immediately follows that 
\begin{align*}
	\mathbb{E}[\hat{S}(t)_{desired}] & = \tr[    \rho'_{desired}(t)   X ] + i \tr[   \rho'_{desired}(t) \,   Y ]\\
	&=
	S(t)_{desired}
\end{align*}

\section{Details of \cref{fig:cliffordt}}

We used the grdidsynth package~\cite{ross2014optimal} to 
optimally synthesize Clifford + T sequences that approximate $R_z(\theta)$ to a precision
at least $\epsilon \leq 10^{-2.4}$ as we described in the main text.
150 digits of precision was needed to achieve machine precision in the solution vector.

\section{Details of \cref{fig:opt_cont}}

The cost function $| \tr[ U(d, )  R^\dagger_x(\pi/2)]|$ was maximised
via a gradient descent optimisation of the pulse amplitudes.
The single-qubit transfer matrices needed 400 digits of precision to achieve machine precision in
the exact solution.


%

\end{document}